\documentclass[checkin,showpacs,aps,prl,noshowkeys]{revtex4}
\usepackage{threeparttable}
\usepackage{amsmath}
\usepackage{lipsum}
\usepackage{multirow}
\usepackage{booktabs}
\usepackage{graphicx,color}
\usepackage{graphicx}
\usepackage{subfigure}
\usepackage{amsmath}

\newcommand{\ba}{\begin{eqnarray}}
\newcommand{\ea}{\end{eqnarray}}

\newcommand{\be}{\begin{equation}}
\newcommand{\ee}{\end{equation}}

\newcommand{\et}{{\it et al. }}

\definecolor{pink}{rgb}{1,0.18,1.0}
\def\prl{{ Phys. Rev. Lett. }}

\def\prb{{ Phys. Rev. B }}

\def\sci{{ Science }}

\def\nm{{Nature Mater. }}

\def\jpcm{{J. Phys.: Condens. Matter }}

\def\nc{{Nat. Commun. }}

\begin{document}

\title{Negative Poisson's ratios in few-layer orthorhombic arsenic from
  first-principles calculations}

\author{Jianwei Han}
\affiliation{Key Laboratory for Magnetism and Magnetic Materials of
 the Ministry of Education, Lanzhou University, Lanzhou 730000, China}

\author{Jiafeng Xie}
\affiliation{Key Laboratory for Magnetism and Magnetic Materials of
 the Ministry of Education, Lanzhou University, Lanzhou 730000, China}

\author{Z. Y. Zhang}
\affiliation{Key Laboratory for Magnetism and Magnetic Materials of
 the Ministry of Education, Lanzhou University, Lanzhou 730000, China}

\author{D. Z. Yang}
\affiliation{Key Laboratory for Magnetism and Magnetic Materials of
 the Ministry of Education, Lanzhou University, Lanzhou 730000, China}

\author{M. S. Si$^{*}$}
\affiliation{Key Laboratory for Magnetism and Magnetic Materials of
 the Ministry of Education, Lanzhou University, Lanzhou 730000, China}

\author{D. S. Xue$^{\dagger}$}
\affiliation{Key Laboratory for Magnetism and Magnetic Materials of
 the Ministry of Education, Lanzhou University, Lanzhou 730000, China}
\date{\today}

\begin{abstract} A material exhibiting a negative Poisson's ratio is
  always one of the leading topics in materials science, which is due
  to the potential applications in those special areas such as
  defence and medicine. In this letter, we demonstrate a new material, few-layer
  orthorhombic arsenic, also possesses the negative Poisson's ratio. For
  monolayer arsenic, the negative Poisson's ratio is predicted to be
  around -0.09, originated from the hinge-like structure within the
  single layer of arsenic. When the layer increases, the negative Poisson's ratio
  becomes more negative and finally approaches the limit at four-layer,
  which is very close to the bulk's value of -0.12. The underlying mechanism is
  proposed for this layer-dependent negative Poisson's ratio, where
  the internal bond lengths as well as the normal Poisson's ratio
  within layer play a key role. The study like ours sheds new light on
  the physics of negative Poisson's ratio in those hinge-like
  nano-materials.  
\end{abstract}



\maketitle

Virtually all known materials would undergo a lateral contraction when
stretched longitudinally and vice versa, which is protected by the
conservation of volume under elastic loading. This gives a positive
Poisson's ratio ($\nu$). In isotropic materials, the positive Poisson's
ratio is theoretically in the range from 0 (Cork) to 0.5
(Rubber) \cite{gibson}. If a lateral dimension expands during stretching or vice
versa, the exhibited  Poisson's ratio is negative and the associated
material is termed auxetic \cite{lakes}.  Much intense interest in
this counterintuitive feature stems from the pioneering discoveries
that critical fluids, re-entrant polymer foams, colloidal crystals,
laser-cooled crystals, unscreened metals, $\alpha$-cristobalite, A7
structure elements arsenic and bismuth, and laminates are found to be
auxetic  \cite{greaves,haeri,gunton,milton}.  
Such an unusual mechanical property of  $\nu < 0$ makes a material
possible for applications in those special areas. For example,
compression in one direction results in a shrink not an expansion in
the transverse direction, demonstrating the mechanism of bulletproof
vests in national defence \cite{liu}; The opposite situation$-$an
expansion responds to a stretch$-$is the manifestation of artificial
limbs in medicine \cite{evans}.  This sparks a surge in research activity in novel
materials with negative Poisson's ratios.

In 2008, Hall and his colleagues found that negative Poisson's ratios
can appear in low-dimensional carbon sheets when multiwalled nanotubes are
introduced \cite{hall}. This is a big step towards the design of
nanostructured composites, artificial muscles, gaskets, and chemical
and mechanical sensors. Naturally, it raises a question whether the
negative Poisson's ratio can emerge in other low-dimensional
materials. In fact, it is difficult to measure this negative Poisson's
ratio in experiments as the observation of such a ratio is spurious
\cite{grimvall}. But, one can always predict it theoretically
\cite{keskar}. For example, monolayer of black phosphorus (BP) is
reported to exhibit a negative Poisson's ratio through using {\it ab
  initio} method \cite{jiang}.  This negative Poisson's ratio is
ascribed to the occurrence of hinge-like structure within a single
layer of BP. Like BP, arsenic also possesses the hinge-like
structures in the orthorhombic phase. More importantly, two groups
independently demonstrate the thermal stability of its few-layer forms
in this orthorhombic phase \cite{zhiya, kamal}. Their bandgaps  are
both layer- and strain-dependent and the gap values are around 1
eV. More surprising is that the carrier mobility can approach as high
as several thousand 
cm$^{2}$V$^{-1}$s$^{-1}$ in these few-layer arsenic. All these make
few-layer arsenic promising for future semiconducting
applications. However, up to now, no study has been focused on the
mechanical properties, in particular the negative Poisson's ratio. Thus, at
present, it is timely to check the possible existence of negative
Poisson's ratios in few-layer orthorhombic arsenic.

In this Letter, the negative Poisson's ratio is reported to occur in
few-layer arsenic through using first-principles calculations, which
stems from the hinge-like structure within one single layer of arsenic. The negative
Poisson's ratio is about -0.09 at monolayer. When layer increases, the
negative Poisson's ratio becomes more negative and finally approaches
the bulk's value of -0.12 at four-layer. This layer-dependent negative
Poisson's ratio is demonstrated to closely follow the internal bond
lengths, in particular the one perpendicular to the within layer
direction, which indeed increase with layer. To better understand the
underlying mechanism, a rigid mechanical model is proposed, shedding
new light on the negative Poisson's ratio in those hinge-like vdW
nano-materials.

First-principles calculations in this work are performed within the
framework of density functional theory, as
implemented in the {\tt SIESTA} code \cite{siesta}. We have used the
generalized gradient approximation in the form of Perdew,  Burke, and
Ernzerhof functional \cite{pbe}. The effect of van der  Waals (vdW)
interaction is taken into account by using the empirical correction
scheme proposed by Cooper \cite{cooper}.
Only the valence electrons are considered in the
calculation, with the core being replaced by norm-conserving scalar
relativistic pseudopotentials \cite{pseudo} factorized in the
Kleinman-Bylander form \cite{kleinman}. We have used a split-valence
double-$\zeta$ basis set including polarization orbitals with an
energy shift of 100 meV for all atoms \cite{cutoff}. The convergence
is achieved when the difference of the total energies between two
consecutive ionic steps is less than 10$^{-5}$ eV and the maximum
force allowed on each atom is set to be 0.01 eV/{\AA}.

\begin{figure}
\includegraphics[width=16cm]{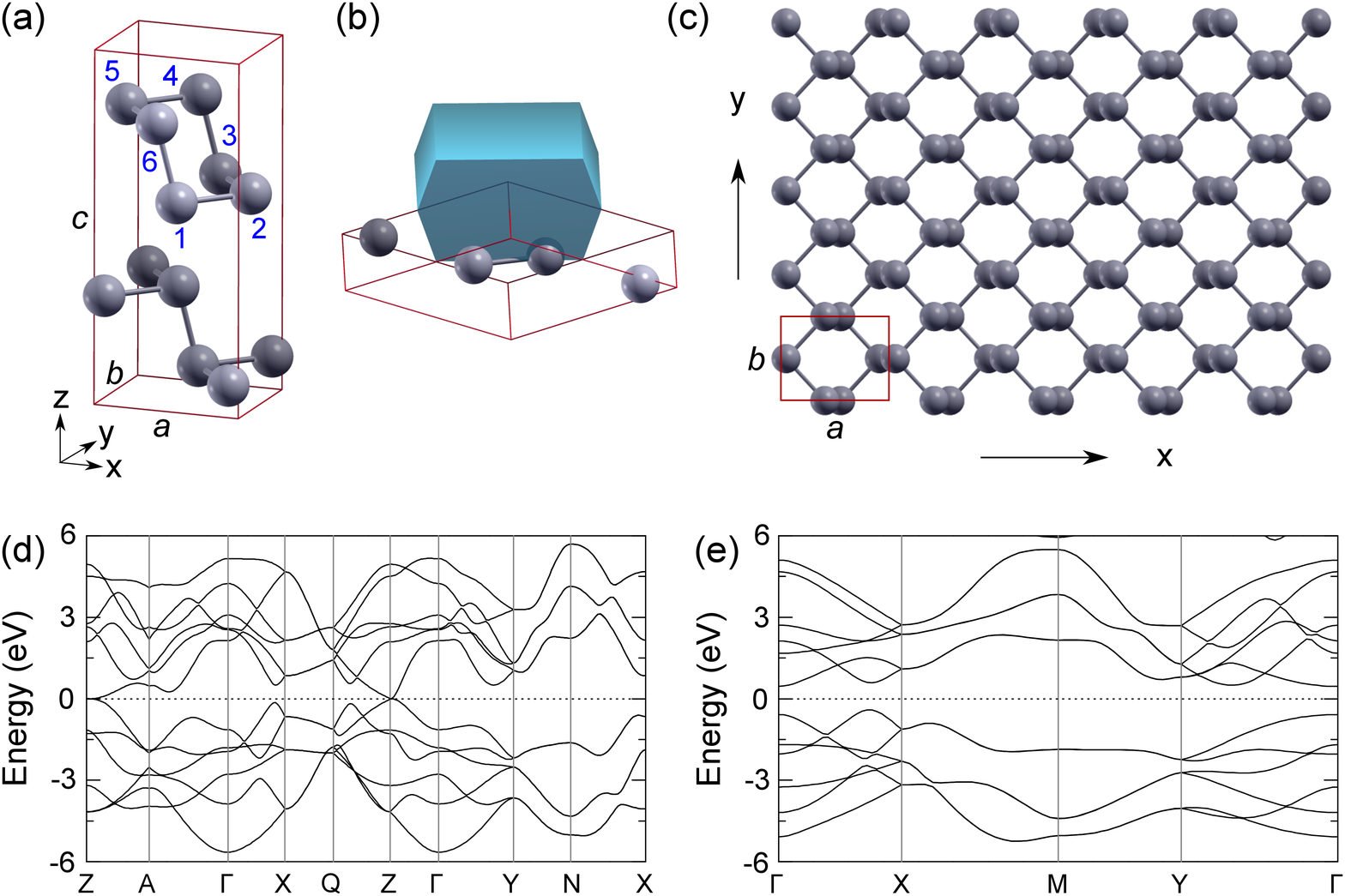}
\caption{(color online). (a) Conventional unit cell of orthorhombic
  arsenic with lattice constants $a$, $b$, and $c$ and internal
  atoms 1-6. (b) Primitive unit cell and its
  Wigner-Seitz cell (blue shaded configuration).  (c) Top view of
  monolayer arsenic with rectangle showing the unit cell. Band
  structures of (d) bulk and (e) monolayer. }
\label{fig1}
\end{figure}

We start our work from the orthorhombic bulk arsenic with space group
{\it Cmca}, which is the same as BP. The conventional
unit cell includes eight atoms, as shown in Fig. \ref{fig1}(a). Each
As atom within a single layer is covalently bonded with three As
atoms, forming a puckered graphene-like hexagonal structure.  This
puckered structure is also called a hinge-like structure, which
consists of two orthogonal hinges (atoms 456 and 612). This sets up
the basis for those exotic properties in few-layer arsenic.
The lattice constants are optimized to $a$ = 4.70 
{\AA}, $b$ = 3.77 {\AA}, and $c$ = 11.11 {\AA}, generating the internal
parameters $r_{12}$ = 2.58 {\AA},   $r_{34}$ = 2.56 {\AA},
$\theta_{123}$ = 94.09$^{\rm o}$, and $\theta_{234}$ = 99.05$^{\rm
  o}$. All these agree well with experimental and theoretical results
\cite{exp,zhiya}.  A primitive unit cell and its Wigner-Seitz cell are
illustrated in Fig. \ref{fig1}(b). It is comprised of four
inequivalent atoms, which is only one half of the bulk case. In
principle, monolayer of arsenic (arsenene), as shown in
Fig. \ref{fig1}(c), can be obtained through exfoliating its bulk
counterpart. Once it is formed, some changes occur.  The corresponding
lattice constants $a$ and $b$ are changed to be 4.73 {\AA} and 3.71
{\AA}, respectively. In comparison with its bulk phase, $a$ increases
while $b$ decreases. This gives a direct effect on the internal parameters:
bond lengths $r_{12}$ and $r_{34}$ are decreased to 2.53 {\AA} and
2.50 {\AA}, while the bond angle $\theta_{123}$ is increased to
94.54$^{\rm o}$ and leaves $\theta_{234}$ = 100.36$^{\rm o}$ nearly
unchanged.  As the thermal stability of a material is important for
real device applications, here we confirm it from the
cohesive energy.  The calculated cohesive energy of orthorhombic bulk
arsenic is 2.90 eV/atom, which is smaller by 0.15 eV/atom than that of
the A7 structure arsenic \cite{kittel}. This coincides with the fact
that the A7 structure is the favored phase in experiment. However, when
they are isolated into monolayers, they become equally stable as the
cohesive energy difference is only $\sim$0.01 eV/atom.  This means
monolayer of orthorhombic arsenic is possible in experiment.

The calculated band structure of bulk phase is displayed in
Fig. \ref{fig1}(d). It clearly shows that the valence band maximum (VBM)
  and the conduction band minimum (CBM) locate at the same crystal
  momentum Z point, demonstrating a direct bandgap semiconductor with nearly zero gap
  value.  However, when it goes into monolayer,  it behaves as an indirect
  bandgap semiconductor, as shown in Fig. \ref{fig1}(e). This is
  because that VBM appears along the $\Gamma$-X line and close to the
  X point, while CBM locates at the $\Gamma$ point. The obtained
  bandgap is about 1 eV. The underlying mechanism of the bandgap
  transition from direct (bulk) to indirect (monolayer) is dominated
  by the mutual competition of the two interlayer bondings $r_{12}$ and
  $r_{34}$, see also the discussion in our previous work
  \cite{zhiya}. In the following, we pay attention on the mechanical
  properties of few-layer arsenic.

\begin{figure}
\includegraphics[width=16cm]{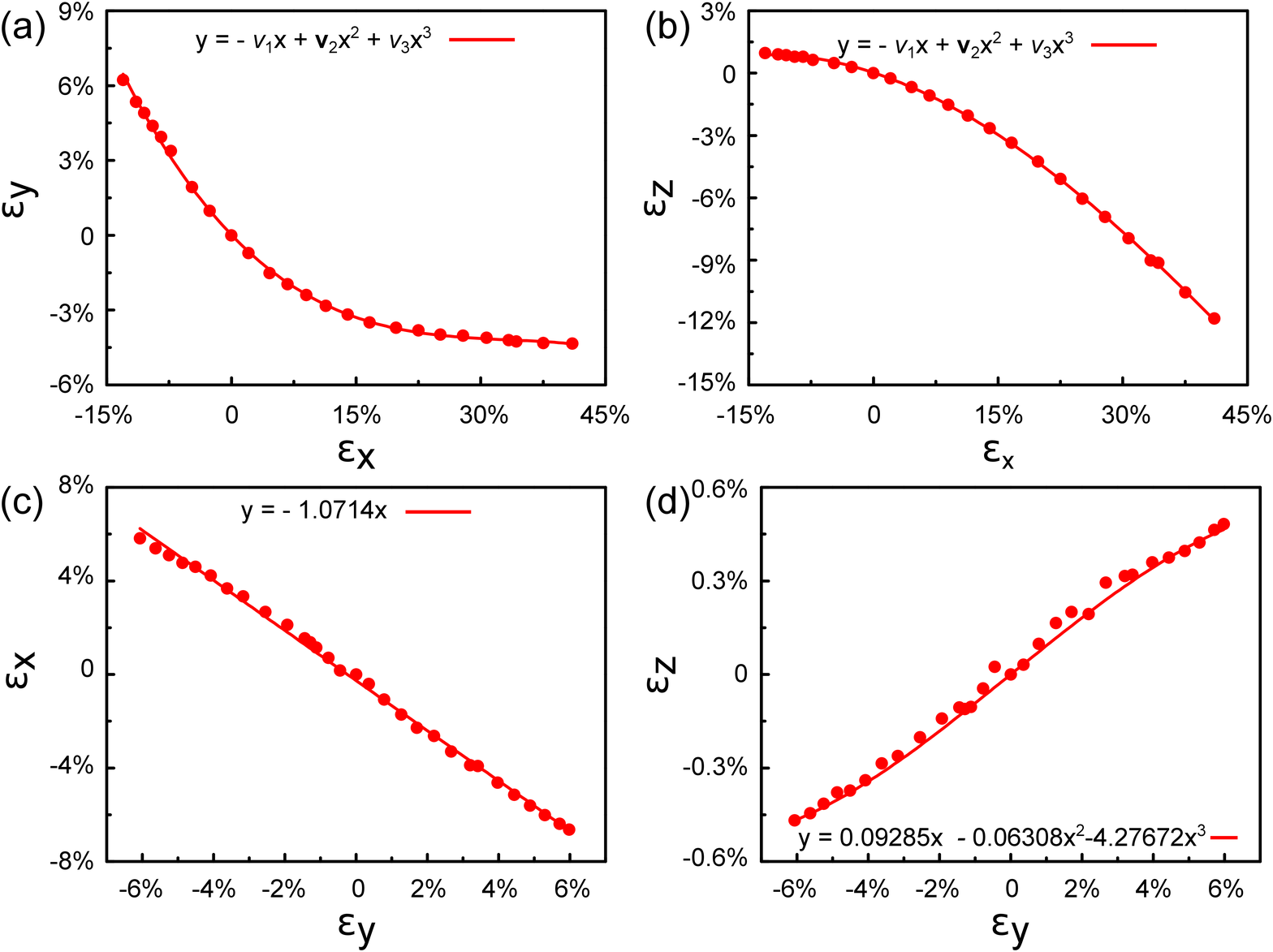}
\caption{(a) $\varepsilon_{y}$ versus
  $\varepsilon_{x}$. The solid circles are simulated data and the line is
fitted by function $y=-\nu_{1}x+\nu_{2}x^{2}+\nu_{3}x^{3}$, with
$\nu_{1}$ = 0.35 as the linear Poisson ratio, $\nu_{2}$ = 1.00 and
$\nu_{3}$ = -1.00.  (b) $\varepsilon_{z}$ versus 
$\varepsilon_{x}$. The fitted linear Poisson ratio is obtained to
$\nu_{1}$ = 0.13, as well as $\nu_{2}$ = -0.48 and $\nu_{3}$ =
0.23. (c)  $\varepsilon_{x}$ versus  $\varepsilon_{y}$. Data are
fitted to function $y=-\nu x$, with $\nu$ = 1.07 as the linear Poisson
ratio. (d) $\varepsilon_{z}$ versus  $\varepsilon_{y}$. Data are
fitted to function $y=-\nu_{1}x+\nu_{3}x^{3}$, with
the negative Poisson ratio $\nu=\nu_{1}$ = -0.093.}
\label{fig2}
\end{figure}

When deformation is applied along the {\it x} direction, the
responding strain in the {\it y} direction is shown in
Fig. \ref{fig2}(a). Note that the positive (or negative)
$\varepsilon$ means a tensile (or compressive) strain.  The
obtained data (solid circles) behaves as a strongly nonlinear feature, which is well
fitted by function of $y =-\nu_{1}x+\nu_{2}x^{2}+\nu_{3}x^{3}$. The
linear parameter $\nu_{1}$ is fitted to be 0.35 and  can be regarded as
the linear Poisson's ratio. Similarly, we can obtain the linear Poisson's
ratio in the {\it z} direction, as shown in Fig. \ref{fig2}(b). The
corresponding linear Poisson ratio is $\nu=\nu_{1}$ = 0.13, which is
nearly twice that in the {\it y} direction. This
means that {\it z} direction is harder than {\it y} direction when
they respond to the strain applied along the {\it x} direction. 
It should be noticed that in the hinge-like structure of arsenic (see
Fig. \ref{fig1}(a)), the deformation of {\it z} direction
is dominated by the bond length $r_{34}$ and the bond angle
$\theta_{234}$, while that of {\it y} direction needs the changes of 
$r_{12}$ and $\theta_{123}$. Perturbed by the same external field,
such as the strain in our case, the bond angle is more easily affected
than that of bond length. Thus, we can infer that the bond angle
$\theta_{123}$, in comparison with $\theta_{234}$, are largely changed
when the deformation is applied along the {\it x} direction. This
coincides with our first-principle calculations.

If the deformation is applied along the {\it y} direction, the
situation dramatically changes. In Fig. \ref{fig2}(c), $\varepsilon_{x}$
linearly depends on $\varepsilon_{y}$. The fitted Poisson's ratio is
$\nu =1.07$, which is much larger than the Poisson's ratios when the
deformation is applied along the {\it x} direction. This is the direct
manifestation of anisotropic feature in mechanical properties, which
is also the origin of anisotropic transport reported in our previous
work \cite{zhiya}.  Up to now, all the obtained Poisson's ratios are
normal and positive. However,  when we check the strain along the {\it
  z} direction responding to the {\it y}-direction loading, the law
behaved as in Fig. \ref{fig2}(d) is opposite. $\varepsilon_{z}$
increases with $\varepsilon_{y}$, leading to an unusual Poisson's
ratio, namely $\nu < 0$. The fitted linear Poisson's ratio is $\nu$ =
-0.093. Its magnitude is very huge if one remembers the negative
Poisson's ratio of around -0.027 in monolayer of BP \cite{jiang}. This
makes monolayer arsenic more suitable for applications in those
special areas such as aerospace and defence where large negative
Poisson's ratios are requested. For monolayer arsenic, the internal
bond lengths are larger than those of 2.42 {\AA} and 2.38 {\AA} in
monolayer BP \cite{jiang}. In principle, under the same loading,
larger bonding would bear a stronger deformation. This explains why 
monolayer arsenic holds a larger (more) negative Poisson's ratio,
being consistent with that $\nu$ mostly increases with atomic number
{\it Z} \cite{greaves}.

\begin{figure}
\includegraphics[width=16cm]{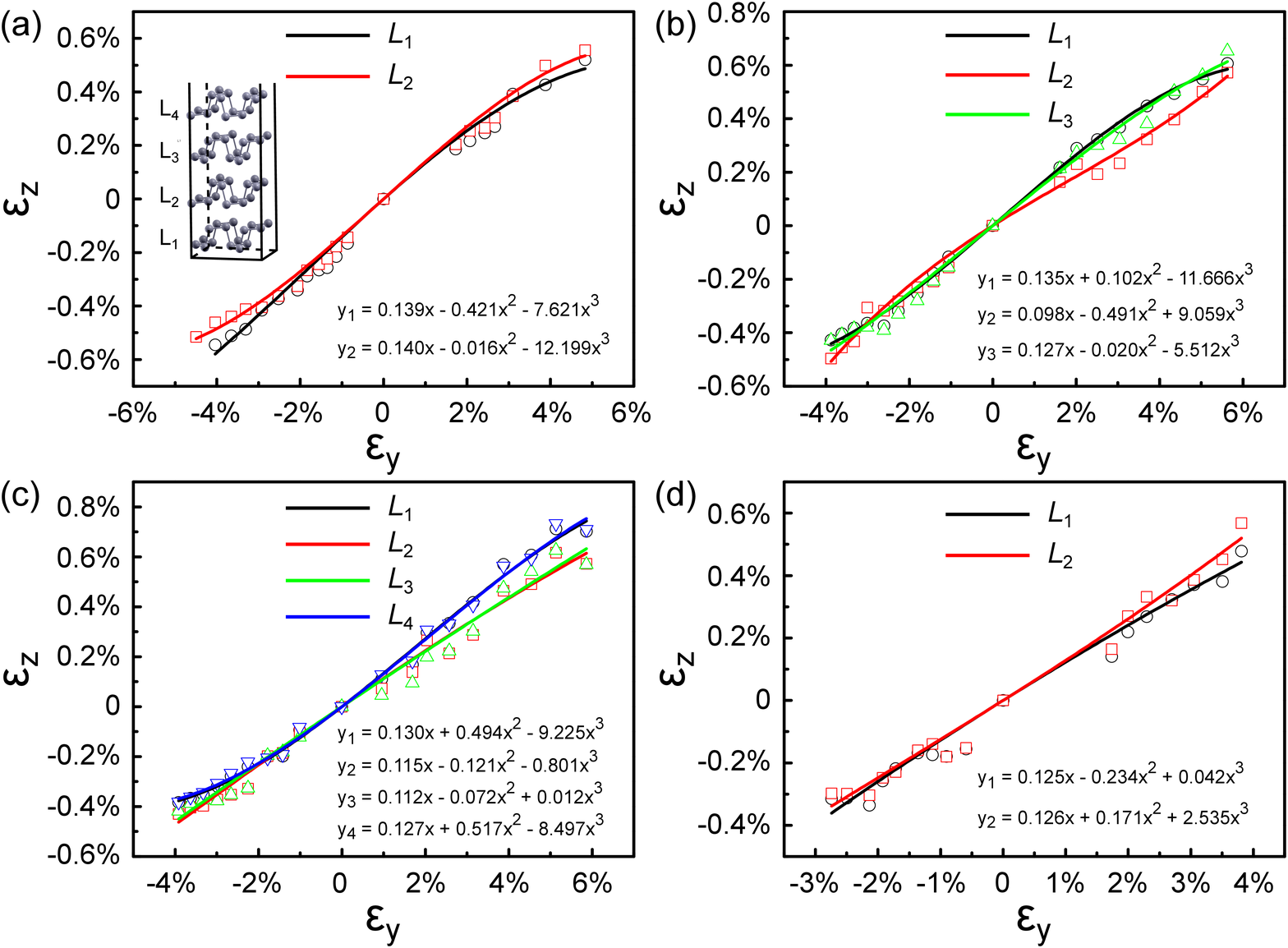}
\caption{(color online). (a)  $\varepsilon_{z}$ versus
  $\varepsilon_{y}$ for bilayer arsenic. Inset shows a four-layer
  arsenic with layers labeled as $L_{1}-L_{4}$ from bottom to
  top. The fitted functions for each layer are given as
  well. $\varepsilon_{z}$ versus $\varepsilon_{y}$ for (b) trilayer,
  (c) four-layer and (d) bulk arsenic. Other denotes are the same as in (a).}
\label{fig3}
\end{figure}

It is well known that layer stacking is emerging as a new degree of
freedom to tune the electronic properties of vdW hetero-materials
\cite{geim}. However, its impact on the negative Poisson's ratio is
still missing. In the following, we will demonstrate how it affects the
negative Poisson's ratios in few-layer arsenic. Once an additional
layer is added, the negative Poisson's ratio is changed accordingly
as the effect of layer stacking is involved, as displayed in
Fig. \ref{fig3}(a). In comparison with monolayer, the
nonlinear feature of $\varepsilon_{z}$ versus  $\varepsilon_{y}$ is
significantly enhanced for bilayer. This is represented by
the larger fitted parameters of cubic terms, which are at least twice
that of monolayer.  In the vicinity of zero, $L_{1}$ and $L_{2}$ are
nearly identical. When the strain is beyond around $\pm$4\%, $L_{1}$
and $L_{2}$ largely separate from each other. This is because the two
layers are no longer equivalent in bilayer due to the $AB$
stacking. The separated feature in the stress-strain curves under larger
strain reflects the asymmetric structure of layers, confirming the
validity of our simulations. The linear Poisson's ratio is fitted to
be about -0.14, which is increased by $\sim$50\% compared with that of
monolayer. Based on the above discussion, the larger the bonding
length, the larger the negative Poisson's ratio. Compared to the bond
length $r_{12}$,  the bond length $r_{34}$ is almost parallel to the {\it z}
direction, which is directly related to the negative Poisson's
ratio. Due to the interlayer vdW interaction introduced by an
additional layer, $r_{34}$ is increased to be 2.54 {\AA}, resulting in
a larger negative Poisson's ratio in bilayer.

In the case of trilayer, as shown in Fig. \ref{fig3}(b), the
stress-strain curves $L_{1}$ and $L_{3}$ are identical, leaving the curve
$L_{2}$ deviated from $L_{1}$ and $L_{3}$ in the whole stress
range. All these coincide with the symmetry of trilayer where layers
$L_{1}$ and $L_{3}$ are symmetrically equivalent, but not for layer
$L_{2}$. This leads to two negative Poisson's ratios. One is about
-0.13 for the layers $L_{1}$ and $L_{3}$. The other one is about
-0.098 for the layer $L_{2}$, which is smaller than those of $L_{1}$
and $L_{3}$. This is because the length $r_{34}$ of layer $L_{2}$ is 2.51
{\AA}, which is smaller than the value  of 2.55 {\AA} for layers
$L_{1}$ and $L_{3}$. When any more layer is added, the internal bond
lengths in particular $r_{34}$ will be not changed largely. For
example, in the case of four-layer, the bond lengths $r_{34}$ are 2.55
{\AA} and 2.54 {\AA} for $L_{1}$ (or $L_{4}$)  and $L_{2}$ (or
$L_{3}$), which are very close to the value of 2.55 {\AA} in the bulk
arsenic. The obtained negative Poisson ratio of four-layer is about
-0.121 (averaged value), as shown in Fig. \ref{fig3}(c). This value
approaches the bulk's value of -0.125, as shown in
Fig. \ref{fig3}(d). This means that the quantum effect on the negative
Poisson's ratios is limited within four layers.

\begin{figure}
\includegraphics[width=16cm]{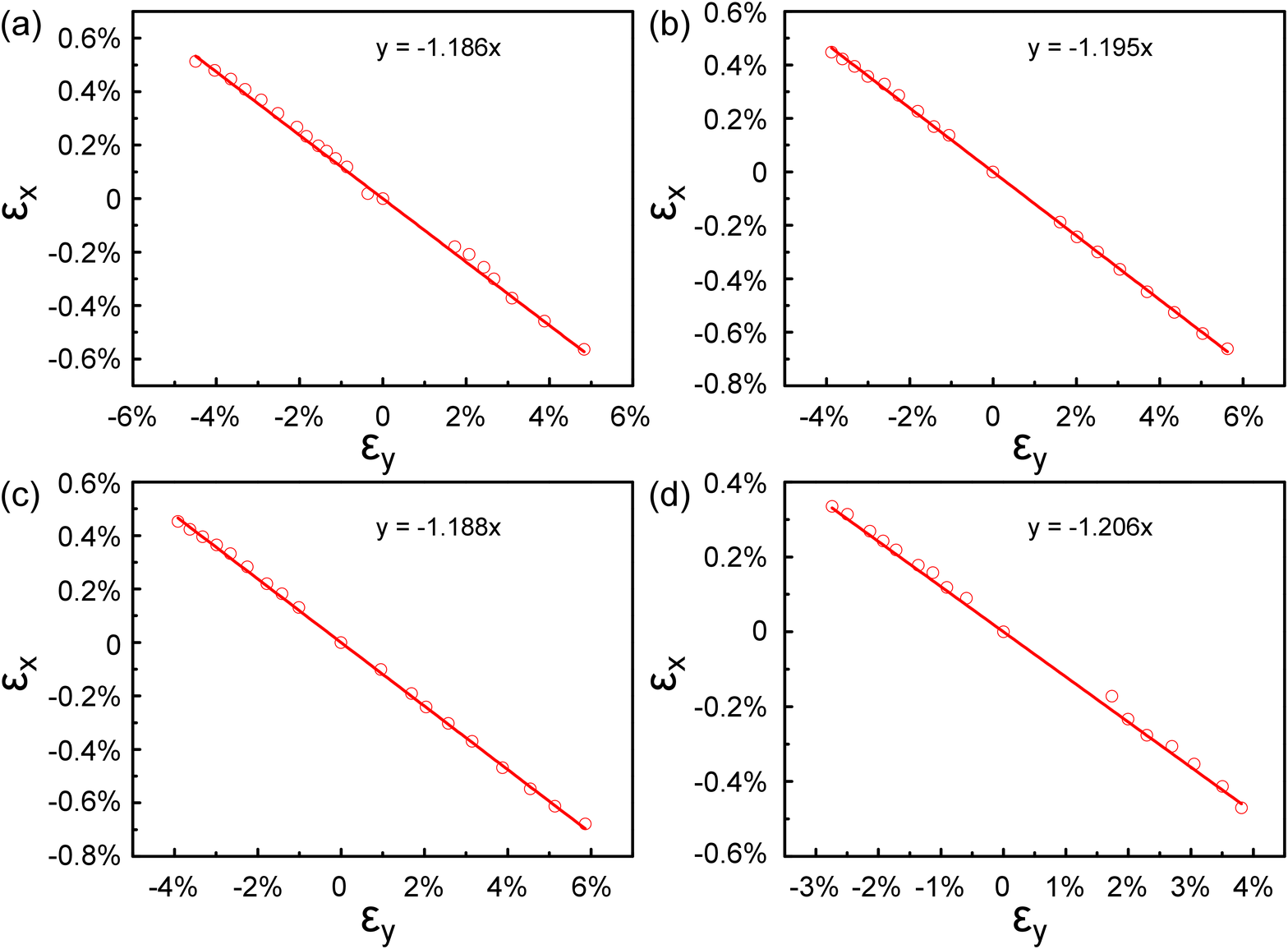}
\caption{ $\varepsilon_{x}$ versus $\varepsilon_{y}$ for (a) bilayer, (b) trilayer,
  (c) four-layer and (d) bulk arsenic. The fitted functions are also given.}
\label{fig4}
\end{figure}

To further understand the layer stacking on the negative Poisson's
ratios, we demonstrate it from the intrinsic puckered structure
combined with the associated stress-strain curves
$\varepsilon_{x}$ versus $\varepsilon_{y}$. As pointed out above, the
hinge-like structure is the origin of the negative Poisson's ratios in
few-layer arsenic. The reason is as follows.  When few-layer arsenic
is stretched in the {\it y} direction, the layer contracts in the {\it
  x} direction, protected by the normal Poisson's ratios, as shown in
Figs. \ref{fig1}(c) and \ref{fig4}(a)-\ref{fig4}(d). In other
words, atoms 3 and 4 as well as atoms 1 and 6 move inward along the
{\it x} direction when stretched in the {\it y} direction (see
Fig. \ref{fig1}(a)). This directly causes the bond angles
$\theta_{234}$ and $\theta_{216}$ smaller compared to the initial
values. Taking into account all the values of $\theta_{234}$ and
$\theta_{216}$ (with/without strain) being larger than 90$^{\rm o}$,
the layer thickness along the {\it z} direction (the projected
distance of $r_{34}$ or $r_{16}$) is increased, leading to a negative
Poisson's ratio. Therefore, the bond angles $\theta_{234}$ and $\theta_{216}$ being
larger than 90$^{\rm o}$ is another necessary condition for the
negative Poisson's ratios in few-layer arsenic. This is also true for
BP as the bond angle $\theta_{234}$ is 97.64$^{\rm o}$ in BP
\cite{jiang}. If a material possesses the bond angles $\theta_{234}$
and $\theta_{216}$ being smaller than 90$^{\rm o}$, the negative
Poisson's ratio will disappear. Future researches can test this
prediction.

Due to the layer stacking, the normal Poisson's ratios are increased
from  1.07 at monolayer (Fig. \ref{fig1}(c)) to 1.19 at four-layer
(Fig. \ref{fig4}(c)). This implies that the decrease of $\theta_{234}$
or $\theta_{216}$ under the same stretching in the {\it y} direction
is enhanced as layer increases. As a result, the negative Poisson's
ratio increases with layer. For four-layer arsenic, the normal
Poisson's ratio is very close to the bulk's value (Figs. \ref{fig4}(c)
and \ref{fig4}(d)).  This is the reason why the negative Poisson's
ratio can  approach the bulk's limit at four-layer.  This tells us that
the effect of layer stacking goes into the negative Poisson's ratio in
the {\it y} direction which is indeed through the normal Poisson's
ratio in the {\it x} direction.

In conclusion, the negative Poisson's ratio is for the first time
reported in few-layer arsenic through using first-principles
calculations.  The magnitude of negative Poisson's ratio is about 0.1
at monolayer, which is about five times of magnitude larger than that
of monolayer BP, suggesting extended applications in those special
areas such as defence and medicine. When layer increases, the
negative Poisson's ratio become large (more negative). The limited
value of around -0.12 is predicted at four-layer, which
is very close to the bulk's value. The underlying mechanism is
demonstrated based on the layer-dependent internal bond length $r_{34}$ and the
normal Poisson's ratio of $\varepsilon_{x}$ versus
$\varepsilon_{y}$. The study like ours sheds new light on the
layer-dependent negative Poisson's ratio in those hinge-like vdW
nano-materials, which will evolve into an active field.

This work was supported by the National Basic Research Program of
China under Grant No. 2012CB933101 and  the National Science
Foundation under Grant No. 51372107, No. 11104122 and No. 51202099. 
This work was also supported by the National Science Foundation for
Fostering Talents in Basic Research of the National Natural Science
Foundation of China. We also acknowledge this work as done on Lanzhou
University's high-performance computer Fermi.

$^{*}$Email: sims@lzu.edu.cn

$^{\dagger}$Email: xueds@lzu.edu.cn

\clearpage

\clearpage

\clearpage

\clearpage

\begin{thebibliography}{99}

\bibitem{gibson} L. J. Gibson, K. E. Easterling, and M. F. Ashby,
  Proc. R. Soc. London Ser. A {\bf 377}, 99 (1981).

\bibitem{lakes} R. S. Lakes, \sci {\bf 235}, 1038 (1987).

\bibitem{greaves} G. N. Greaves, A. L. Greer, R. S. Lakes, and
  T. Rouxel, \nm {\bf 10}, 823 (2011).

\bibitem{haeri} A. Y. Haeri, D. J. Weidner, and J. B. Parise, Science {\bf
  257}, 650 (1992).

\bibitem{gunton} D. J. Gunton and G. A. Saunders, J. Mater. Sci. {\bf
  7}, 1061 (1972).

\bibitem{milton} G. Milton, J. Mech. Phys. Solids {\bf 40}, 1105 (1992).

\bibitem{liu} Q. Liu, {\it Literature Review: Materials with Negative
  Poisson's Ratios and Potential Applications to Aerospace and
  Defence}, (DSTO, Defence Science and Technology Organisation, 2006).

\bibitem{evans} K. E. Evans and K. L. Alderson, Adv. Mater. {\bf 12},
  617 (2000).

\bibitem{hall} L. J. Hall, V. R. Coluci, D. S. Galv\~ao, M. E. Kozlov,
  M. Zhang, S. O. Dantas, and R. H. Baughman, \sci {\bf 320}, 504
  (2008).

\bibitem{grimvall} G. Grimvall, {\it Thermophysical Properties of
  Materials}, (North-Holland, Amsterdam, 1986).

\bibitem{keskar} N. R. Keskar and J. R. Chelikowsky, Nature {\bf 358},
  222 (1992). 

\bibitem{jiang} J.-W. Jiang and H. S. Park, \nc {\bf 5}, 4727 (2014). 

\bibitem{zhiya} Z. Y. Zhang \et arXiv:1411.3165.

\bibitem{kamal} C. Kamal and M. Ezawa, arXiv:1410.5166.  

\bibitem{siesta} J. M. Soler, E. Artacho, J. Gale, A. Garc\'ia, J. Junquera,
P. Ordej\'on, and D. S\'anchez-Portal, \jpcm {\bf 14}, 2745 (2002).

\bibitem{pbe} J. P. Perdew, K. Burke, and M. Ernzerhof, \prl {\bf 77},
  3865 (1996).

\bibitem{cooper} V. R. Cooper, \prb {\bf 81}, 161104(R) (2010).

\bibitem{pseudo} N. Troullier and J. L. Martins, \prb {\bf 43}, 1993
  (1991).

\bibitem{kleinman} L. Kleinman and D. M. Bylander, \prl {\bf 48}, 1425
  (1982). 

\bibitem{cutoff} E. Artacho, D. S\'anchez-Portal, P. Ordej\'on, A. Garc\'ia, and J. M.
Soler, Phys. Status Solidi B {\bf 215}, 809 (1999).

\bibitem{exp} P. M. Smith, A. J. Leadbetter, and A. J. Apling,
  Philos. Mag. B {\bf 31}, 57 (1975).

\bibitem{kittel} C. Kittel, {\it Introduction to Solid State Physics},
  eighth ed. (Wiley, Hoboken, NJ, 2004).

\bibitem{geim} A. K. Geim and I. V. Grigorieva, Nature {\bf 499}, 419 (2013). 

\end{thebibliography}
\end{document}